# Nitrate removal from water using electrostatic regeneration of functionalized adsorbent

James W. Palko[a,b,†], Diego I. Oyarzun[a,†], Byunghang Ha[a], Michael Stadermann[c], Juan G. Santiago[a,*]

[a] *Department of Mechanical Engineering, Stanford University, Stanford, CA 94305, USA*
[b] *Department of Mechanical Engineering, University of California, Merced, CA 95340, USA*
[c] *Lawrence Livermore National Laboratory, Livermore, CA 94550, USA*
[†] contributed equally

Nitrate is an important pollutant in drinking water worldwide, and a number of methods exist for the removal of nitrate from water including ion exchange and reverse osmosis. However, these approaches suffer from a variety of disadvantages including requirements for supply and disposal of brine used for regeneration in ion exchange and low water recovery ratio for reverse osmosis. Here, we demonstrate the use of high surface area activated carbon electrodes functionalized with moieties having high affinity for adsorption of nitrate from aqueous solution, such as those used in ion exchange. Adsorption of surfactant molecules having a quaternary amine ionic group to the activated carbon surfaces provides functionalization of the surfaces without complex chemistries. The functionalized electrodes have adsorption capacities of about 80 mg $NaNO_3$ per gram of activated carbon material. Unlike a traditional ion exchanger, the functionalized surfaces can be repeatedly regenerated by the application of an electrostatic potential which displaces the bound $NO_3^-$ while leaving an excess of electronic charge on the electrode. The cell is completed by an inert counter electrode where Faradaic reactions occur during regeneration. Up to approximately 40% of the initial capacity of the electrode is accessible following electrical regeneration.

## 1. Introduction

Nitrate is a pollutant of major and rapidly increasing concern in drinking water worldwide. High nitrate levels in groundwater are associated with changes in the balance of the nitrogen cycle associated with intensive agriculture. The growing use of fertilizers and worldwide increase in agricultural intensity continue to increase the concern over this pollutant.[1]–[3] Nitrate itself is rather benign, but it has the potential for reduction to toxic nitrite in the human digestive system. Infants are considered highly sensitive to high nitrate levels, with significant risk of developing methemoglobinemia, a potentially fatal condition reducing the oxygen carrying capacity of the blood. Livestock are likewise at risk from poisoning from high nitrate levels. Nitrates in food and drinking water are also implicated in the generation of carcinogenic nitrosamines. [4]–[7]

Due to the often diffuse nature of nitrate contamination, the most compatible methods for nitrate treatment should be distributed, point-of-entry, or point-of-use systems.[8] One of the most popular methods of treatment for nitrate removal in distributed settings is ion exchange.[9], [10] Ion exchange is a commonly applied technique to treat water with a range of ionic contaminants (e.g. nitrate, perchlorate, alkaline earth ions). In ion exchange, charged groups (e.g. anionic quaternary amines) attached to a resin are initially weakly bound to counter charged ions (e.g. $Cl^-$). When exposed to a solution containing ions with a higher chemical affinity (e.g. $NO_3^-$), these ions displace the lower affinity ions and are removed from the solution. Lower affinity ions are "exchanged" for higher affinity ions in solution based on the difference in their bound-to-free equilibrium constants. The relative affinities are influenced by the size, bonding, and electronic structure of the ions and surface groups and may be difficult to predict.[11] Resins for nitrate removal often use standard anion exchange quaternary amine moieties such as trimethyl amine (type I) or dimethylethanol amine (type II) bound to divinylbenzene pendant groups of the polymer matrix.[12], [13] These groups show higher affinity for sulfate ions and so their capacity for nitrate is reduced for feed solutions high in sulfate. Nitrate specific resins have been developed, which employ moieties such as triethyl- and tributylamine.[9],

---

[*] Corresponding author. Email: juan.santiago@stanford.edu

[14] Ion exchange is highly effective and widely applied for a variety of water treatment needs such as hardness and nitrate removal. However, the standard technique suffers from several drawbacks. When all lower affinity ions on the resin have been exchanged for higher affinity ions, the resin is exhausted and must be regenerated or replaced. When regenerated in place, a brine (e.g. NaCl) is applied to the resin bed at sufficient concentrations (e.g. 1 M) to drive the equilibrium toward the state with a majority of the active sites again occupied by weakly bound ions. The regeneration process thus requires significant quantities of concentrated brine, and this poses a number of problems in the implementation of ion exchange systems. The requirements for regular brine tank filling and material cost are significant impediments to application in point-of-use and point-of-entry systems. Although point-of-entry systems are a key segment of treatment equipment, they have particular difficulty with disposal of brine during regeneration, as such installations typically rely on septic systems for waste water disposal. Septic systems may be susceptible to clogging when exposed to highly saline waste streams due to the salting out of weakly soluble oily species.[15] Reverse osmosis systems are also effective for nitrate removal from drinking water.[16], [17] However, they suffer from high costs and low recovery ratios.[18] Biological methods are highly effective for nitrate removal from waste water, but are not currently deployed for drinking water.[19], [20] A number of adsorbents are also capable of passive removal of nitrate from water.[21]

Capacitive deionization (CDI) is an alternative method for removal of ionic contaminants from water.[22]–[24] CDI uses electrostatic interactions to adsorb ionic contaminants from solution onto high surface area electrodes leaving behind purified water which is then flushed from the cell. The electrodes are then discharged to release the contaminants into a concentrated waste stream, which is flushed from the cell with additional feed water to complete the cycle. CDI cells can be used without ion selective membranes[22] or with such membranes in an effort to improve charge efficiency.[25], [26] Further, CDI can be implemented such that ions are adsorbed due to the application of an applied potential difference to electrodes ("traditional" CDI), or ions can be adsorbed without applied potentials by the action of chemical charges on electrode surfaces. In the latter case, ions are removed from the CDI cell by applying a potential—a process termed inverted CDI.[27] Inverted CDI uses native surface charges for the adsorption action, but the role of chemical affinity has not been well understood. Relevant to the current work, the application of CDI to trapping of nitrate has been limited to CDI cells operated in a "traditional mode" (i.e. nitrate ions trapped under applied fields) with carbon electrodes alone [28]–[30] as well as electrodes incorporating anionic selective membranes [31], including membranes with specific affinity for nitrate.[32]–[35] Electrochemical oxidation and reduction has also been applied for the in-situ, transient formation of ionic surface groups with high affinity for ionic solutes on functionalized electrodes as in the electrically switched ion exchange method (ESIX).[36]–[38] ESIX has been applied to the removal of nitrate using conductive polymer active electrodes.[39], [40]

In this study, we leverage the established nitrate affinity of trimethyl quaternary amines to drive nitrate removal from solution via adsorption onto functionalized high surface area activated carbon.[41] This adsorption is performed under no applied potential to the electrode. As in inverted CDI, we then remove the adsorbed ions by application of a potential to the conductive activated carbon substrate. In contrast to inverted CDI, the cell is completed by an inert titanium electrode that passes current during regeneration of the active electrode via Faradaic reactions. A key element of this design is the prevention of readsorption of the expelled nitrate from the active electrode during regeneration.

## 2. Materials and methods
### 2.1. Electrode material
We used commercially available PACMM activated carbon material (Material Methods LLC., Irvine, CA) as the active porous carbon electrode. This material has been previously applied and characterized in CDI.[42]–[46] The electrode has a thickness of ~300 μm.

### 2.2. Surfactant treatment of electrode material
We functionalized the active electrode with trimethyl quaternary amine moieties by adsorption of the common cationic surfactant cetrimonium bromide (CTAB) on the surfaces of the activated carbon (**Figure 1**b). CTAB has been shown to be adsorbed on activated carbon and dramatically increase its capacity to passively adsorb high affinity ions such as perchlorate,[47]–[50] as well as provide surface charges for CDI.[51] We soaked the activated carbon electrodes in aqueous solutions of 10 mM CTAB and 10 mM NaCl overnight with roughly 1.25 g of electrode material per liter of solution. The nonpolar alkane segment of the CTAB molecular adsorbs readily to the AC surface and leaves the quaternary amine head group exposed for later interaction with anions in solution.

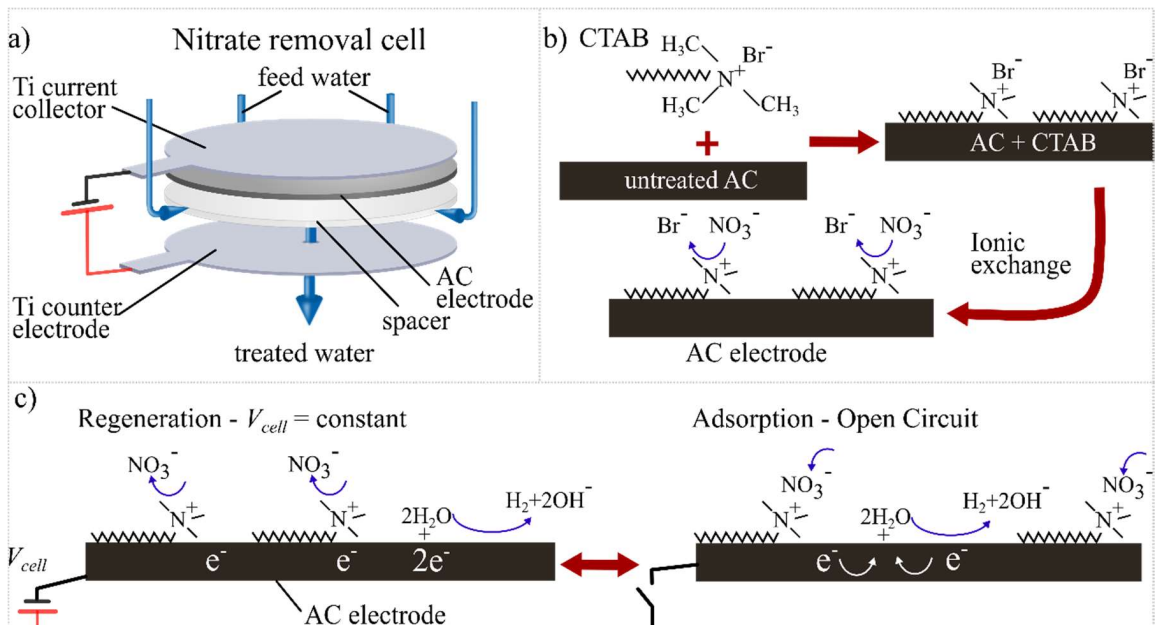

**Figure 1:** a) Schematic of radial flow-between nitrate treatment cell with CTAB functionalized AC electrode and titanium Faradaic counter electrode. Arrows represent flow paths. b) Schematic of CTAB treatment of AC and the initial ionic exchange process in which $NO_3^-$ displaces $Br^-$ due to the higher affinity of the quaternary amine functional group for the former. c) Schematic of the cyclical electrical regeneration and passive adsorption in functionalized electrode. Potential, $V_{cell}$, is applied to the electrode (left) which drives off nitrate, and simultaneously creates Faradaic reactions at the counter electrode and some smaller Faraday current at the AC electrode. Following regeneration, the AC electrode is electrically isolated (open circuit) and passive adsorption of nitrate occurs in the electrode along with the dissipation of the stored charge via Faradaic reactions on the active electrode.

### 2.2.1. Nitrate concentration measurement

We quantified nitrate concentration in solution via optical absorption using UV absorbance in a spectrophotometer (Agilent Cary 6000i UV/Vis/NIR).[52] We measured optical absorption at either 205 nm or 225 nm depending on the presence of bromide ions in solution. The shorter wavelength provided somewhat higher sensitivity to the $NO_3^-$ ion but suffers from significant absorption from $Br^-$ as well. Therefore, we used 225 nm for $NO_3^-$ concentration measurements for the initial exposure of the treated electrode to $NO_3^-$ during which $Br^-$ is expelled.[53] After the initial adsorption, we used adsorption at 205 nm for measurement of nitrate concentration. For the majority of measurements, we collected fractions of the effluent over 30 min intervals into separate samples contained in 15 mL conical plastic tubes using a fraction collector with an effluent flow rate of 0.43 mL/min (13 mL/sample). For short time response measurements, we flowed the effluent directly through the spectrophotometer cuvette for online measurement. Additional details of the nitrate measurement procedure are available in the Supplementary Information (SI) of this paper.

### 2.3. Cell design

Figure 1a shows a schematic of the nitrate treatment cell. The cell uses a radial inflow geometry. Feed water is supplied to the outer periphery of the circular active and counter electrodes, and flows between the two electrodes toward the center of the cell through a woven plastic mesh spacer (300μm thick). The treated water then exits the cell through a hole in the center of the counter electrode, which is formed from titanium (140 μm thick) and operates through Faradaic reactions on its surface. A titanium current collector (50 μm thick, Grade 2) also presses against the active electrode to form electrical contact. A peristaltic pump provides continuous flow of feed water to the cell, and treated effluent is either collected using a fraction collector for analysis or sent directly through a flow cuvette for UV adsorption measurement. The active area of the cell is 3.5 cm in diameter, and the active electrode has a mass of 0.15 g.

## 2.4. Adsorption/regeneration cycle

Immediately following CTAB treatment, the active electrode acts as a traditional ion exchanger (**Figure 1**b). However, in contrast to traditional regeneration methods using high concentration of low affinity ions, here we regenerate the active electrode by the application of a negative potential, relative to the counter electrode (**Figure 1**c), which displaces the $NO_3^-$ ions and leaves behind excess electrical charge in the electrode that balances the positive amine groups on the surface. Following regeneration, the active electrode again acts as an adsorbent with no connection to the counter electrode. As nitrate ions are bound by the amine groups, electronic charge is dissipated from the electrode via Faradaic reactions at its surface. (See **Figure 4** and discussion below).

## 3. Results and discussion
### 3.1. $NO_3^-$ adsorption by functionalized activated carbon saturated with halide ions

Immediately following functionalization of the active electrode, the quaternary ammonium groups are saturated with a mixture of $Br^-$ and $Cl^-$ ions. When exposed to solutions containing ions with higher affinity, such as $NO_3^-$, the higher affinity anions displace surface bound ions of lower affinity (as in a traditional ion exchange resin) (**Figure 1**b). **Figure 2** shows the time response of the effluent from the cell for an electrode with no applied potential which was freshly functionalized and exposed to an input stream containing 200 ppm $NaNO_3$ and with a flow rate of 2.9 mL/min/g-electrode (0.43 mL/min for the cell). Effluent $NaNO_3$ concentration dropped to 45 ppm for 1 h then slowly rose until it approached the input concentration after ~12 h. As shown in the inset of **Figure 2**, the active electrode was saturated with $NO_3^-$ by a cumulative adsorption of 83 mg $NO_3^-$/g-electrode. For comparison, commercial ion exchange resins show a capacity of up to ~150 mg $NO_3^-$/g-electrode.[9] PACMM material that has not been treated with CTAB shows a total $NO_3$ capacity of less than 5 mg $NO_3^-$/g electrode.

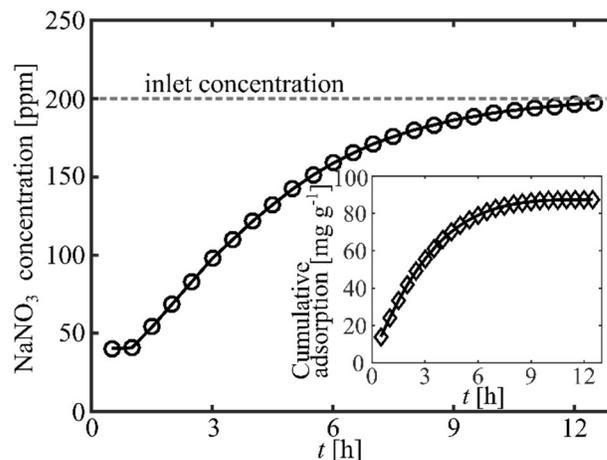

**Figure 2:** Adsorption of $NO_3^-$ by freshly CTAB treated AC electrode. (main) Effluent $NaNO_3$ concentration is shown versus time for 200 ppm $NaNO_3$ inlet concentration. Flow rate was 2.9 mL/min/g-electrode (0.43 mL/min for full cell). Symbols are experimental data points corresponding to times of collection for samples with volume 13 mL each. (inset) Cumulative adsorption on electrode vs. time in mass of $NaNO_3$ normalized by mass of electrode. Total adsorption is 83 mg $NO_3^-$/g-electrode.

### 3.2. Electrical regeneration of functionalized AC

In contrast with ion exchange resins, our functionalized activated carbon electrode can be electrically regenerated to restore adsorption activity. **Figure 3** shows a series of selected regeneration and adsorption cycles (numbered) for the active electrode. (Cycles 6-9 are discussed later, and not shown in **Figure 3**.) The input stream was again maintained with a $NaNO_3$ concentration of 200 ppm and flow rate of 2.9 mL/min/g-electrode (0.43 mL/min for full cell). **Figure 3**a shows the effluent concentration for regeneration using a constant voltage of 3 V applied between the active and counter electrodes for 4 h followed by adsorption at open circuit. Application of negative potential to the active electrode drives off bound $NO_3^-$ anions. Current corresponding to the nitrate ion release (as well as Faradaic reactions on the active electrode) flows to the counter electrode. As the counter electrode has minimal capacitance, Faradaic reactions at the counter electrode accommodate this current. The regeneration voltage of 3 V exceeds the electrolysis threshold for water. Therefore, we expect Faradaic reactions to occur on the active electrode, as well, in parallel with the expulsion of nitrate ions.

Following regeneration, the quaternary amine groups of the CTAB molecules are balanced (at least in part) by electronic charge in the electrode, as seen by the retention of a negative voltage on the active electrode with respect to the counter electrode potential. **Figures 3**b and c show the voltage between the active (-) and counter (+) electrodes and the current leaving the active electrode, respectively, during each regeneration/adsorption cycle. The time axis in **Figure 3** indicates only total active time during cycles. Inactive time between cycles is excluded for clarity of presentation, but the active electrode is exposed to the input nitrate stream at 200 ppm concentration during this time as well.

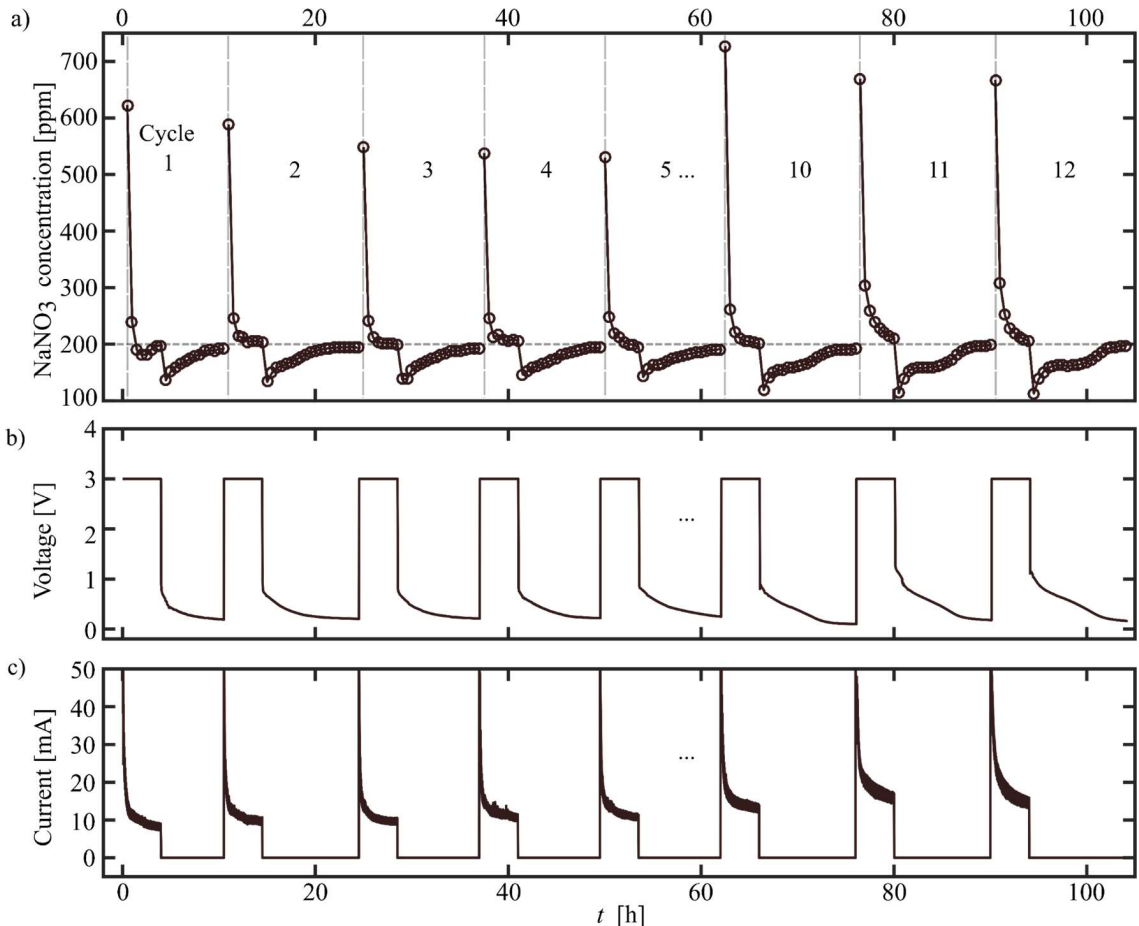

**Figure 3:** Cyclic adsorption of nitrate using electrical regeneration. Time series showing a) Effluent NaNO$_3$ concentration vs. time for input concentration of 200 ppm NaNO$_3$, b) voltage, and c) current versus time for fixed flow rate (Q = 0.43mL/min). For cycles 1 through 5 and 10 through 12, we adsorbed at open circuit, while in cycles 6-7 (shown in SI) we adsorbed at short circuit. The time axis has been abridged to include only active cell times. The length of inactive time between each consecutive cycle are given in the SI.

During adsorption, NO$_3^-$ ions are removed from solution and electronic charge is dissipated from the electrode (**Figure 1**c). This results in a gradual decrease in cell potential during adsorption (**Figure 3**b). The adsorption cycles shown in **Figure 3** occur with an open circuit between the active and counter electrodes. Consequently, the electronic charge in the electrode is dissipated by Faradaic reactions on the electrode itself (**Figure 1**c).

**Figure 4** shows the correspondence of nitrate removal from solution and electronic charge dissipation from the electrode. **Figure 4**a gives time series for nitrate removal rate (represented by the difference in input and effluent NaNO$_3$ concentration at fixed flow rate) and charge dissipation rate of the capacitive electrode (calculated as the time rate of change of potential for the capacitive electrode) during adsorption for each of the cycles shown in **Figure 3**. Adsorption for these cycles occurs at open circuit, so no current flows through the external circuit and the titanium counter electrode essentially acts as a reference electrode. **Figure 4**b shows the correlation between nitrate removal rate (as measured in effluent) and electrode potential discharge rate. Cycles 6-7 (see **Figure S1**) were conducted using a short circuit between the active and counter electrodes. Effluent concentration in these cycles shows similar behavior to that seen in **Figure 1**, indicating dominance of Faradaic reactions on the active electrode.

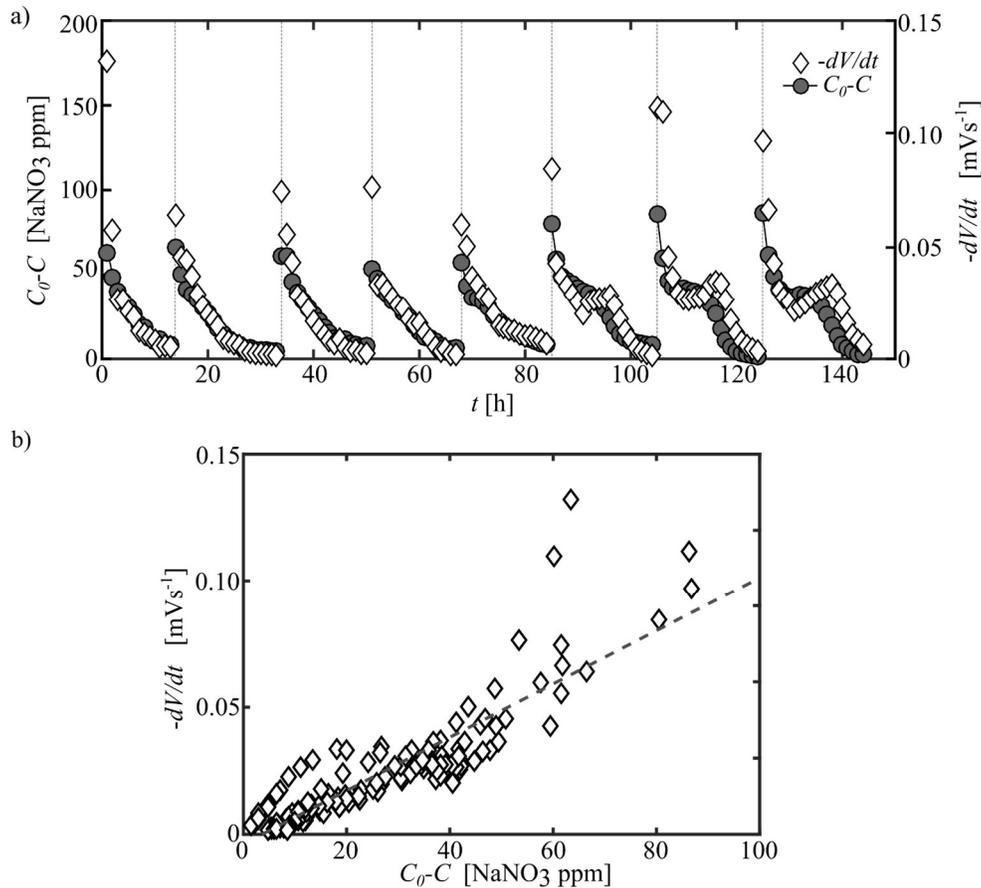

**Figure 4:** a) Adsorption of $NO_3^-$ (left ordinate) by electrically regenerated electrodes versus time (as deduced from effluent stream), and (right ordinate) time derivative of porous electrode voltage during the adsorption cycles versus time. Symbols indicate time of collection for samples with volume 13 ml. b) Correlation plot of nitrate removal versus time derivative of porous electrode voltage during the adsorption cycles at open circuit. Rate of adsorption correlates strongly with discharge of electrode ($dV/dt$) indicating anion adsorption is balanced by removal of electronic charge from electrode by Faradaic reactions.

**Table 1** gives the cumulative $NaNO_3$ adsorption and expulsion for each cycle as well as the maximum and minimum effluent concentrations averaged over 30 min during regeneration and adsorption, respectively. The cumulative mass of $NaNO_3$ released during regeneration per mass of electrode material is 33 mg/g during the first two cycles, but falls with additional cycles. Comparison of the regenerated quantity of $NaNO_3$ with the charge transferred to the active electrode during regeneration (**Figure 3**c) indicates that a larger majority (e.g. >95%) of the cell current during regeneration corresponds to Faradaic reactions on the active electrode. The cumulative adsorbed mass measured per cycle is consistently below the released mass, averaging ~20 mg/g for the first 7 cycles. We hypothesize that this difference is due to additional adsorption that occurs while the cell is offline, and perhaps, chemical changes in the active or counter electrodes. Evaporation during collection or errors in flowrate may also contribute.

We noted dramatic reductions in effluent nitrate concentration during purge for cycles 8 and 9, and minimal absorption capacity after purge. Following cycle 9, we replaced the counter electrode with a fresh titanium sheet but retained the previously used active electrode. This replacement resulted in a significant improvement in both regeneration and adsorption performance, as seen from **Figure 3**a. After counter electrode replacement, the cumulative $NaNO_3$ mass released during regeneration rises to ~45 mg/g, and the cumulative adsorption increases to ~36 mg/g. The transient response of effluent concentration and cell voltage also show a notable difference following replacement of the counter electrode, displaying a plateau in effluent concentration and an inflection point in electrode potential discharge rate. The mechanisms for these differences in regeneration and adsorption behavior associated with changes in the counter electrode are not clear. Titanium is commonly used as an anode in electrochemical systems due to the extreme corrosion resistance imparted by the surface oxide coating. [54] We tested the possibility of excessive oxidation on the surface by polishing both the titanium current collector and counter electrode with abrasive, but observed no effect. We hypothesize that impurities in the electrode may play a significant role in the Faradaic reactions occurring at the counter electrode during purge, and these may leach from the electrode during operation, reducing its effectiveness as they are depleted. We also believe improved contact between the active electrode material and current collector following reassembly of the cell may play a role in the behavior observed. A more robust electrode material such as platinum or platinized titanium may offer superior durability. As we describe below, we hope as part of future work to replace the Faradaic electrode in our cell with a porous capacitive counter electrode to minimize Faradaic currents and improve power efficiency.

**Table 1:** Cumulative adsorbed and expelled NaNO₃ per unit mass of electrode, as well as maximum and minimum NaNO₃ concentration for each cycle from **Figure 3**a. Input concentration was 200 ppm NaNO₃ and flow rate was 0.43 mL min$^{-1}$.

| Cycle | Adsorption [mg g$^{-1}$] | min. $C$ [ppm] | Regeneration [mg g$^{-1}$] | max. $C$ [ppm] |
|---|---|---|---|---|
| 1  | 17.1 | 136 | 32.5 | 623 |
| 2  | 20.0 | 133 | 32.7 | 588 |
| 3  | 21.6 | 138 | 28.4 | 548 |
| 4  | 20.1 | 146 | 30.3 | 538 |
| 5  | 20.7 | 142 | 28.2 | 530 |
| 6  | 22.1 | 141 | 28.2 | 509 |
| 7  | 18.6 | 144 | 23.2 | 465 |
| 10 | 34.4 | 119 | 42.1 | 727 |
| 11 | 35.8 | 113 | 47.8 | 669 |
| 12 | 37.6 | 113 | 44.9 | 667 |

The adsorption capacity accessible through electrical regeneration was consistently lower than for the initially treated electrode. **Figure 5**a compares the adsorption behavior for the active electrode immediately after CTAB treatment and for the cyclically operating electrode operated with multiple electrical regenerations. The comparisons show that only a fraction of the initial adsorbance capacity was accessible during normal operation with electrical regeneration.

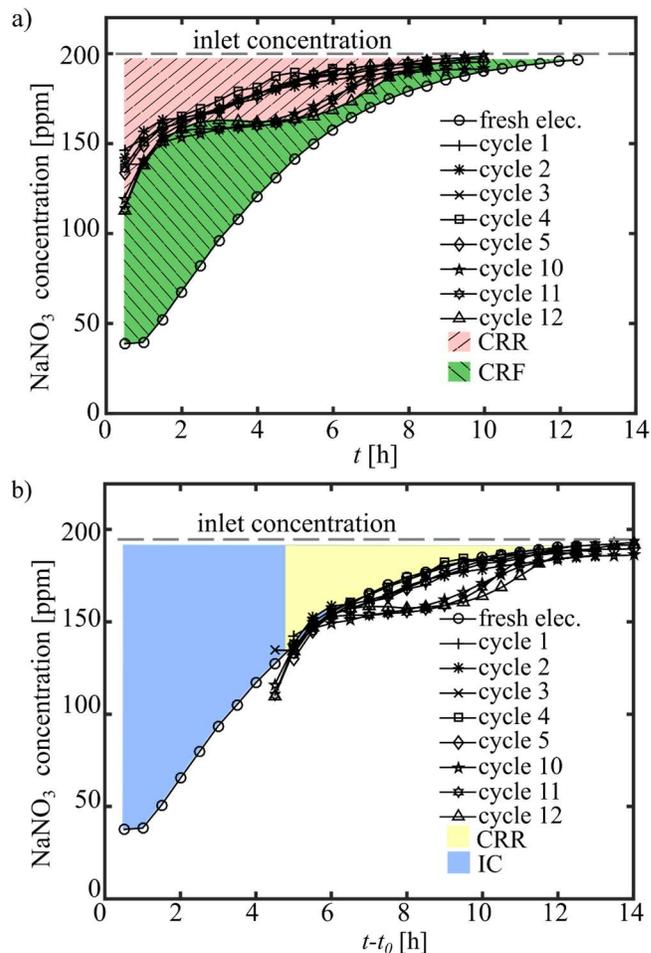

**Figure 5:** a) Adsorption behavior for AC electrode following electrical regeneration in comparison with fresh CTAB treatment. Series show effluent NaNO₃ concentration vs. time for 200 ppm NaNO₃ inlet concentration. Hatched areas represent maximum cumulative NO₃⁻ removal following electrical regeneration (CRR, cumulative removal with electrical regeneration) and initially treated electrode (CRF, cumulative removal fresh electrode). b) Effluent NaNO₃ concentration vs. time with time axis for each series shifted such that electrode saturation times (i.e. time where effluent concentration equals inlet concentration) are coincident. Time shift shows that adsorption rate is strongly correlated to degree of electrode saturation. Area CRR is representative of adsorption following electrical regeneration. Left area corresponds to inaccessible capacitance, IC, for NO₃⁻ adsorption possibly due to electrical limitations. Symbols indicate time of collection for samples with volume 13mL. Flow rate is 0.43 mL/min.

For cycles 1-5 only about 24% of the initial capacity was available. After replacement of the counter electrode (cycles 10-12), about 43% of the capacity was repeatedly accessible.

### 3.3. Possible $NO_2^-$ generation

The significant voltages applied during regeneration provide the potential for electrochemical reduction of $NO_3^-$ to $NO_2^-$. We tested for the presence of nitrite ions in the cell effluent during regeneration with a range of cell voltages from 1 to 8 V. Using a colorimetric test (LaMotte Insta-TEST 2996 Nitrate/Nitrite test strips, LaMotte Co., Chestertown, MA), with a detection limit of ~1.5 ppm $NO_2^-$, we detected no nitrite production in the effluent for regeneration voltages below 4 V. At cell voltages above 4 V nitrite is produced in the effluent stream.

### 3.4. $NO_3^-$ adsorption following electrical regeneration

The plots of **Figure 5** show how the rate of nitrate removal depends strongly on the degree of nitrate saturation in the active electrode. To highlight this dependence, we show in **Figure 5**b the effluent concentrations following electrical regeneration shifted along the time axis (by ~4.5 h). This enables a more direct comparison between the adsorption capacity of regenerated electrode versus the later stages of adsorption for the freshly treated electrode. Interestingly, the electrode displays similar relationships between the rate of nitrate removal and remaining available adsorption capacity following electrical regeneration and initial treatment.

**Figure 6**a shows measurements of nitrate removal rate following electrical regeneration at 5 V as a function of time for three different flow rates. Increased flow rate offers increase in removal rate, although the increase in removal rate is not directly proportional to flow rate. **Figure 6**b shows the trade-off between the increased removal rate of nitrate removal and the increased nitrate concentration of the effluent, as the cell's removal rate cannot keep up with the demands of increased flow rate. We hypothesize that the trends of **Figures 6**a and b are due to a complex coupling between mass transport limitations, adsorption rate kinetics, and the finite capacity of the cell. For example, we hypothesize that the rate of diffusion is strongly influenced by the near electrode surface concentrations of nitrate, which result from a coupling between streamwise advection and transverse diffusion of nitrate. We hope to further explore these coupled effects in a future study.

### 4. Conclusions

We have shown a nitrate removal cell which uses functionalized groups on a high surface area electrode for both passive adsorption of nitrate, and electrical, on-demand regeneration of the adsorber using no regeneration chemicals whatsoever. The electrically regenerated passive adsorber is activated carbon with groups having high affinity for nitrate consisting of quaternary amines (CTAB). We show a capacity of ~80 mg $NaNO_3$/g activated carbon which is comparable to that available from ion exchange resins. The bulk electrical conductivity of the activated carbon substrate allows the application of an electrical potential to expel adsorbed nitrate ions and regenerate the ion exchange groups without application of high concentration brine. The electrode retains about 43% of the initial capacity following electrical regeneration, and the electrode can be repeatedly regenerated. The ability to electrically regenerate an active electrode provides potential advantages in terms of reduced maintenance and waste disposal needs. The most important limitation of the current cell is our use here of a Faradaic counter electrode. This Faradic counter electrode is convenient for demonstration but leads to higher regeneration voltages and larger energy consumption during regeneration. In future systems, we will pursue the use of capacitive counter electrodes employing functionalization with opposing surface charges to prevent adsorption of nitrate during regeneration. Such an approach falls within the scope of inverted CDI operation, but takes advantage of the specific chemical affinity of the surface functionalization for high selectivity of ions removed from solution.

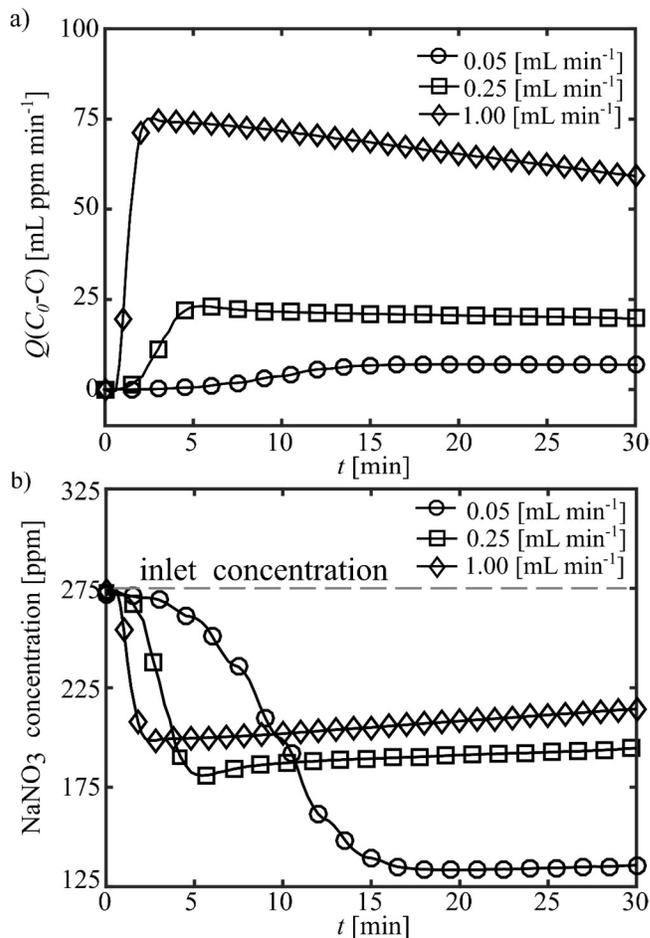

**Figure 6:** Coupling between adsorption rate and flow rate. a) Adsorption rate of $NaNO_3$ versus time, and b) $NaNO_3$ concentration versus time for electrically regenerated AC electrode for various flow rates at a fixed input concentration of 275 ppm $NaNO_3$. Here, electrode was regenerated at 5 V before each adsorption. Strong flow rate dependence of adsorption rate suggests a mass transport limitation.

**Acknowledgements**

This work was funded in part by the TomKat Center for Sustainable Energy at Stanford University. Part of this work was performed at the Stanford Nano Shared Facilities (SNSF), supported by the National Science Foundation under award ECCS-1542152. D.I.O. was supported by a grant from CONICYT / BECAS Chile.

**References**

[1] G. D. Liu, W. L. Wu, and J. Zhang, "Regional differentiation of non-point source pollution of agriculture-derived nitrate nitrogen in groundwater in northern China," *Agric. Ecosyst. Environ.*, vol. 107, no. 2–3, pp. 211–220, 2005.

[2] R. F. Spalding and M. E. Exner, "Occurrence of Nitrate in Groundwater—A Review," *J. Environ. Qual.*, vol. 22, no. 3, p. 392, 1993.

[3] Y. Zhang, F. Li, Q. Zhang, J. Li, and Q. Liu, "Tracing nitrate pollution sources and transformation in surface- and ground-waters using environmental isotopes," *Sci. Total Environ.*, vol. 490, pp. 213–222, 2014.

[4] G. Gulis, M. Czompolyova, and J. R. Cerhan, "An ecologic study of nitrate in municipal drinking water and cancer incidence in Trnava District, Slovakia.," *Environ. Res.*, vol. 88, no. 3, pp. 182–7, 2002.

[5] K. H. Gelberg, L. Church, G. Casey, M. London, D. S. Roerig, J. Boyd, and M. Hill, "Nitrate Levels in Drinking Water in Rural New York State," *Environ. Res.*, vol. 80, no. 1, pp. 34–40, 1999.

[6] A. M. Fan and V. E. Steinberg, "Health implications of nitrate and nitrite in drinking water: an update on methemoglobinemia occurrence and reproductive and developmental toxicity.," *Regul. Toxicol. Pharmacol.*, vol. 23, no. 23, pp. 35–43, 1996.

[7] L. Knobeloch, K. Krenz, H. Anderson, and C. Hovel, "Methemoglobinemia in an infant—Wisconsin," *Morb. Mortal. Wkly. Rep.*, vol. 42, no. 12, pp. 217–219, 1992.

[8] P. Shahbazi, F. Vaezi, A. Hossein Mahvi, K. Naddaffi, and A. Rahmani Reza, "Nitrate Removal from Drinking Water by Point of Use Ion Exchange," *J. Res. Heal. Sci.*, vol. 10, no. 2, pp. 91–97, 2010.

[9] S. Samatya, N. Kabay, Ü. Yüksel, M. Arda, and M. Yüksel, "Removal of nitrate from aqueous solution by nitrate selective ion exchange resins," *React. Funct. Polym.*, vol. 66, no. 11, pp. 1206–1214, 2006.

[10] X. Xu, B. Gao, Y. Zhao, S. Chen, X. Tan, Q. Yue, J. Lin, and Y. Wang, "Nitrate removal from aqueous solution by Arundo donax L. reed based anion exchange resin," *J. Hazard. Mater.*, vol. 203–204, no. 3, pp. 86–92, 2012.

[11] C. E. Harland, Ed., "Ion exchange equilibria," in *Ion Exchange: Theory and Practice*, 2nd ed., The Royal Society of Chemistry, 1994, pp. 90–133.

[12] H. Chen, M. M. Olmstead, R. L. Albright, J. Devenyi, and R. H. Fish, "Metal-Ion-Templated Polymers: Synthesis and Structure ofN-(4-Vinylbenzyl)-1,4,7-Triazacyclononanezinc(II) Complexes, Their Copolymerization with Divinylbenzene, and Metal-Ion Selectivity Studies of the Demetalated Resins—Evidence for a Sandwich Complex in," *Angew. Chemie Int. Ed. English*, vol. 36, no. 6, pp. 642–645, Apr. 1997.

[13] R. E. Barron and J. S. Fritz, "Effect of functional group structure on the selectivity of low capacity anion exchangers for monovalent anions," *J. Chromatogr.*, vol. 284, no. 284, pp. 13–25, 1984.

[14] Y. Zhou, C. D. Shuang, Q. Zhou, M. C. Zhang, P. H. Li, and A. M. Li, "Preparation and application of a novel magnetic anion exchange resin for selective nitrate removal," *Chinese Chem. Lett.*, vol. 23, no. 7, pp. 813–816, 2012.

[15] M. Gross and T. Bounds, "Water Softener Backwash Brine Stresses Household Septic Tanks and Treatment Systems," *Small Flows Mag.*, vol. 8, no. 2, pp. 8–10, 2007.

[16] J. Bohdziewicz, M. Bodzek, and E. Wąsik, "The application of reverse osmosis and nanofiltration to the removal of nitrates from groundwater," *Desalination*, vol. 121, no. 2, pp. 139–147, 1999.

[17] J. J. Schoeman and a. Steyn, "Nitrate removal with reverse osmosis in a rural area in South Africa," *Desalination*, vol. 155, pp. 15–26, 2003.

[18] D. L. Shaffer, N. Y. Yip, J. Gilron, and M. Elimelech, "Seawater desalination for agriculture by integrated forward and reverse osmosis: Improved product water quality for potentially less energy," *J. Memb. Sci.*, vol. 415–416, pp. 1–8, 2012.

[19] C. E. Barrera-Díaz, V. Lugo-Lugo, and B. Bilyeu, "A review of chemical, electrochemical and biological methods for aqueous Cr(VI) reduction," *J. Hazard. Mater.*, vol. 223–224, pp. 1–12, 2012.

[20] E. J. Bouwer and P. B. Crowe, "Biological Processes in Drinking Water Treatment," *Am. Water Work. Assoc.*, vol. 80, no. 9, pp. 82–93, 1988.

[21] A. Bhatnagar and M. Sillanpää, "A review of emerging adsorbents for nitrate removal from water," *Chem. Eng. J.*, vol. 168, no. 2, pp. 493–504, 2011.

[22] J. C. Farmer, D. V Fix, G. V Mack, R. W. Pekala, and J. F. Poco, "Capacitive deionization of NH4ClO4 solutions with carbon aerogel electrodes," *J. Appl. Electrochem.*, vol. 26, no. 10, pp. 1007–1018, 1996.

[23] S. J. Seo, H. Jeon, J. K. Lee, G. Y. Kim, D. Park, H. Nojima, J. Lee, and S. H. Moon, "Investigation on removal of hardness ions by capacitive deionization (CDI) for water softening applications," *Water Res.*, vol. 44, no. 7, pp. 2267–2275, 2010.

[24] Y. Oren, "Capacitive deionization (CDI) for desalination and water treatment — past, present and future (a review)," *Desalination*, vol. 228, no. 1–3, pp. 10–29, Aug. 2008.

[25] H. Li and L. Zou, "Ion-exchange membrane capacitive deionization: A new strategy for brackish water desalination," *Desalination*, vol. 275, no. 1–3, pp. 62–66, 2011.

[26] M. E. Suss, S. Porada, X. Sun, P. M. Biesheuvel, J. Yoon, and V. Presser, "Water desalination via capacitive deionization: what is it and what can we expect from it?," *Energy Environ. Sci.*, vol. 8, no. 8, pp. 2296–2319, 2015.

[27] X. Gao, A. Omosebi, J. Landon, and K. Liu, "Surface charge enhanced carbon electrodes for stable and efficient capacitive deionization using inverted adsorption–desorption behavior," *Energy Environ. Sci.*, vol. 8, no. Cx, p. 897, 2015.

[28] W. Tang, P. Kovalsky, D. He, and T. D. Waite, "Fluoride and nitrate removal from brackish groundwaters by batch-mode capacitive deionization," *Water Res.*, vol. 84, pp. 342–349, 2015.

[29] N. Pugazhenthiran, S. Sen Gupta, A. Prabhath, M. Manikandan, J. R. Swathy, V. K. Raman, and T. Pradeep, "Cellulose Derived Graphenic Fibers for Capacitive Desalination of Brackish Water," *ACS Appl. Mater. Interfaces*, vol. 7, no. 36, pp. 20156–20163, Sep. 2015.

[30] M. Noked, E. Avraham, Y. Bohadana, A. Soffer, and D. Aurbach, "Development of Anion Stereoselective, Activated Carbon Molecular Sieve Electrodes Prepared by Chemical Vapor Deposition," *J. Phys. Chem. C*, vol. 113, no. 17, pp. 7316–7321, Apr. 2009.

[31] R. Broseus, J. Cigana, B. Barbeau, C. Daines-Martinez, and H. Suty, "Removal of total dissolved solids, nitrates and ammonium ions from drinking water using charge-barrier capacitive deionisation," *Desalination*, vol. 249, no. 1, pp. 217–223, Nov. 2009.

[32] J.-H. Yeo and J.-H. Choi, "Enhancement of nitrate removal from a solution of mixed nitrate, chloride and sulfate ions using a nitrate-selective carbon electrode," *Desalination*, vol. 320, pp. 10–16, Jul. 2013.

[33] Y. J. Kim and J. H. Choi, "Selective removal of nitrate ion using a novel composite carbon electrode in capacitive deionization," *Water Res.*, vol. 46, no. 18, pp. 6033–6039, 2012.

[34] Y. J. Kim, J. H. Kim, and J. H. Choi, "Selective removal of nitrate ions by controlling the applied current in membrane capacitive deionization (MCDI)," *J. Memb. Sci.*, vol. 429, pp. 52–57, 2013.

[35] J. S. Koo, N.-S. Kwak, and T. S. Hwang, "Synthesis and properties of an anion-exchange membrane based on vinylbenzyl chloride-styrene-ethyl methacrylate copolymers," *J. Memb. Sci.*, vol. 423, pp. 293–301, Dec. 2012.

[36] M. A. Lilga, R. J. Orth, J. P. H. Sukamto, S. D. Rassat, J. D. Genders, and R. Gopal, "Cesium separation using electrically switched ion exchange," *Sep. Purif. Technol.*, vol. 24, no. 3, pp. 451–466, 2001.

[37] S. D. Rassat, J. H. Sukamto, R. J. Orth, M. A. Lilga, and R. T. Hallen, "Development of an electrically switched ion exchange process for selective ion separations," *Sep. Purif. Technol.*, vol. 15, no. 3, pp. 207–222, 1999.

[38] M. A. Lilga, R. J. Orth, J. P. H. Sukamto, S. M. Haight, and D. T. Schwartz, "Metal ion separations using electrically switched ion exchange," *Sep. Purif. Technol.*, vol. 11, no. 3, pp. 147–158, Jul. 1997.

[39] M. Hojjat Ansari and J. Basiri Parsa, "Removal of nitrate from water by conducting polyaniline via electrically switching ion exchange method in a dual cell reactor: Optimizing and modeling," *Sep. Purif. Technol.*, vol. 169, pp. 158–170, 2016.

[40] M. Hojjat Ansari, J. Basiri Parsa, and J. Arjomandi, "Application of conducting polyaniline, o-anisidine, o-phenetidine and o-chloroaniline in removal of nitrate from water via electrically switching ion exchange: Modeling and optimization using a response surface methodology," *Sep. Purif. Technol.*, vol. 179, pp. 104–117, 2017.

[41] G. Darracq, J. Baron, and M. Joyeux, "Kinetic and isotherm studies on perchlorate sorption by ion-exchange resins in drinking water treatment," *J. Water Process Eng.*, vol. 3, no. C, pp. 123–131, 2014.

[42] A. Hemmatifar, J. W. Palko, M. Stadermann, and J. G. Santiago, "Energy breakdown in capacitive deionization," *Water Res.*, vol. 104, pp. 303–311, 2016.

[43] R. Zhao, P. M. Biesheuvel, and a. van der Wal, "Energy consumption and constant current operation in membrane capacitive deionization," *Energy Environ. Sci.*, vol. 5, no. 11, p. 9520, 2012.

[44] P. M. Biesheuvel, H. V. M. Hamelers, and M. E. Suss, "Theory of Water Desalination by Porous Electrodes with Immobile Chemical Charge," *Colloids Interface Sci. Commun.*, vol. 9, pp. 1–5, Nov. 2015.

[45] J. E. Dykstra, R. Zhao, P. M. Biesheuvel, and A. Van der Wal, "Resistance identification and rational process design in Capacitive Deionization," *Water Res.*, vol. 88, pp. 358–370, 2016.

[46] R. Zhao, M. van Soestbergen, H. H. M. Rijnaarts, a. van der Wal, M. Z. Bazant, and P. M. Biesheuvel, "Time-dependent ion selectivity in capacitive charging of porous electrodes," *J. Colloid Interface Sci.*, vol. 384, no. 1, pp. 38–44, 2012.

[47] S. Y. Lin, W. F. Chen, M. T. Cheng, and Q. Li, "Investigation of factors that affect cationic surfactant loading on activated carbon and perchlorate adsorption," *Colloids Surfaces A Physicochem. Eng. Asp.*, vol. 434, pp. 236–242, 2013.

[48] J. Xu, N. Gao, Y. Deng, M. Sui, and Y. Tang, "Perchlorate removal by granular activated carbon coated with cetyltrimethyl ammonium bromide," *J. Colloid Interface Sci.*, vol. 357, no. 2, pp. 474–479, May 2011.

[49] J. hong Xu, N. yun Gao, Y. Deng, M. hao Sui, and Y. lin Tang, "Perchlorate removal by granular activated

carbon coated with cetyltrimethyl ammonium chloride," *Desalination*, vol. 275, no. 1–3, pp. 87–92, 2011.

[50] R. Mahmudov, C. Chen, and C. P. Huang, "Functionalized activated carbon for the adsorptive removal of perchlorate from water solutions," *Front. Chem. Sci. Eng.*, vol. 9, no. 2, pp. 194–208, 2015.

[51] M. Andelman, "Ionic Group Derivitized Nano Porous Carbon Electrodes for Capacitive Deionization," *J. Mater. Chem. Eng.*, vol. 2, no. March, pp. 16–22, 2014.

[52] A. C. Edwards, P. S. Hooda, and Y. Cook, "Determination of Nitrate in Water Containing Dissolved Organic Carbon by Ultraviolet Spectroscopy," *Int. J. Environ. Anal. Chem.*, vol. 80, no. 1, pp. 49–59, 2001.

[53] A. HIOKI and J. W. MCLAREN, "Direct Determination Method of Nitrate Ions in Seawater by UV-Detection Ion-Chromatography with Hydrochloric Acid/Sodium Chloride Eluent," *AIST Bull. Metrol.*, vol. 7, no. 2, pp. 1–9, 2008.

[54] G. Chen, "Electrochemical technologies in wastewater treatment," *Sep. Purif. Technol.*, vol. 38, no. 1, pp. 11–41, 2004.

# Nitrate removal from water using electrostatic regeneration of functionalized adsorbent

James W. Palko[a,b,†], Diego I. Oyarzun[a,†], Byunghang Ha[a], Michael Stadermann[c], Juan G. Santiago[a,*]

[a] *Department of Mechanical Engineering, Stanford University, Stanford, CA 94305, USA*
[b] *Department of Mechanical Engineering, University of California, Merced, CA 95340, USA*
[c] *Lawrence Livermore National Laboratory, Livermore, CA 94550, USA*
[†] contributed equally

**Abstract**
The supporting information presented here provides descriptions of UV spectrophotometric nitrate measurements, nitrate removal cycles at short circuit, and inactive times between runs for the nitrate removal cell.

**List of Figures and Tables**

## 1. UV spectrophotometric measurements of nitrate concentration

We measured $NO_3^-$ concentrations using a UV/Vis/NIR spectrophotometer (Agilent Cary 6000i) with a short path length (0.2 mm) quartz cuvette. To this end, we generated custom calibration curves for $NO_3^-$ adsorption at wavelengths of 205 and 225 nm. **Figure S1** shows the calibration curves relating solution concentration and absorbance. We used a wavelength of 205 nm for determination of nitrate concentration when other anions were not present in solution (i.e. for cell cycles applying electrical regeneration after the initial ion exchange had completed). $NO_3^-$ shows relatively strong absorption at this wavelength.

During the initial ion exchange of the first adsorption cycle following electrode surfactant treatment we detected the presence of $Br^-$ ions in solution. For such solutions, we used a wavelength of 225 nm for nitrate concentration quantification. Bromide absorbs strongly at shorter wavelengths, but this absorption decays as wavelength increases to ~220 nm.

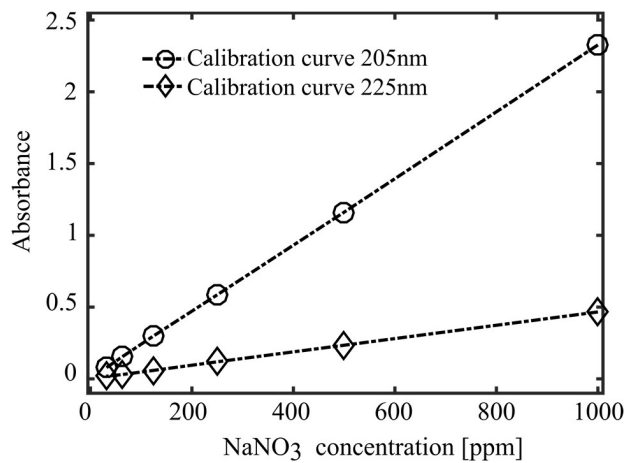

**Figure S 1:** Calibration curves for UV absorbance at 205 and 225 nm vs. $NaNO_3$ concentration for aqueous solutions using a 0.2 mm path length quartz cuvette. Shown are linear regression fits of the form y =x 2.33E- 3 (R = 1.00) for 205nm and y = x 4.674E-4 (R = 0.999) for 225nm.

**Figure S2a** shows the absorption spectra for several concentrations of NaBr. At a wavelength of 225 nm the absorption of $Br^-$ is negligible compared to $NO_3^-$. **Figure S2b** shows the absorption of 200 ppm $NaNO_3$ with varying concentrations of NaBr.

## 2. Nitrate adsorption under short circuit conditions

In addition to the open circuit adsorption cycles following electrical regeneration, two cycles (6 and 7) were conducted using a short circuit between the active and counter electrodes. Effluent concentration, cell voltage, and current are shown versus time in

---

[*] Corresponding author. Email: juan.santiago@stanford.edu

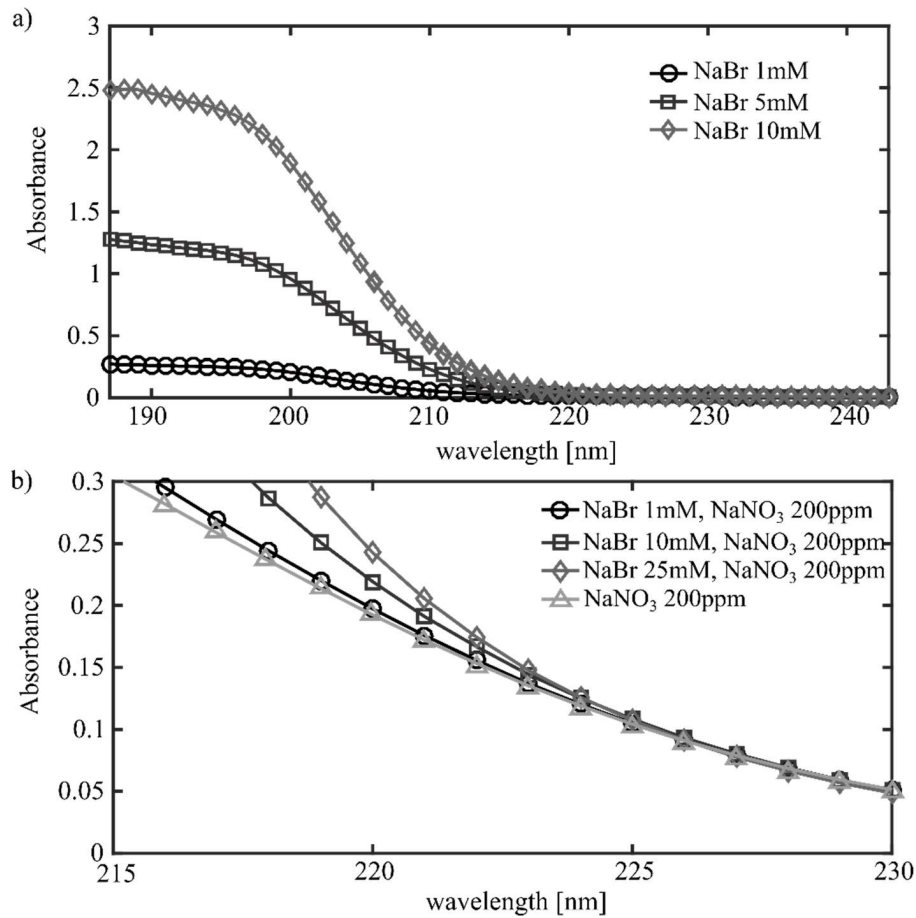

**Figure S 2:** UV Spectrophotometric measurements of anionic concentrations using a quartz cuvette with a 0.2 mm path length. a) Sodium Bromide absorbance vs. wavelength at 1, 5 and 10 mM concentrations. b) Absorbance vs. wavelength of $NaNO_3$ at 200 ppm without additional solutes and in solution with Sodium Bromide (1, 10, and 25 mM). Bromide ions show negligible absorbance at wavelengths greater than 225 nm.

**Figure S3**. In these cycles, we used the same treated activated carbon electrode used for cycles shown in the main text **Figure 3**. Flow rate was fixed to Q = 0.43mL/min. Effluent concentration in these cycles shows similar behavior to that seen in open circuit cycles, indicating dominance of Faradaic reactions on the active electrode during adsorption.

### 3. Cell inactive times

Cumulative times reported in the main text for multiple cycles refer to the times in which the cell is actively adsorbing or regenerating. Inactive times between runs are reported in Table S1. Zero delay between cycles refers to consecutive cycles. Inactive times were used to collect samples and inspect the cell

**Table S 1:** Inactive delay time after each adsorption cycle.

| Cycle | Delay after cycle [h] |
|---|---|
| 1 | 3 |
| 2 | 10 |
| 3 | 1 |
| 4 | 0 |
| 5 | 3 |
| 6 | 0 |
| 7 | 3 |
| 8 | 0 |
| 9 | 10 |
| 10 | 6 |
| 11 | 0 |
| 12 | - |

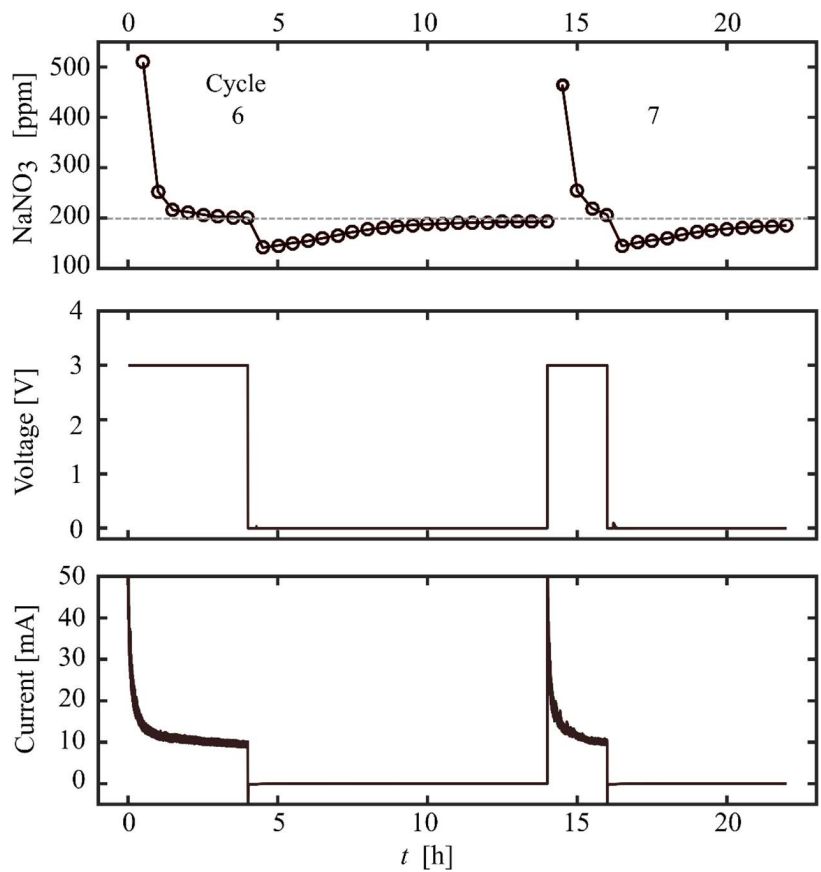

**Figure S 3:** Cyclic adsorption of nitrate with active and counter electrodes shorted following electrical regeneration. Time series showing a) Effluent NaNO$_3$ concentration, b) voltage, and c) current versus time for input concentration of 200 ppm NaNO$_3$ and fixed flow rate of Q = 0.43mL/min. The time axis has been abridged to include only active cell times.